\documentclass[a4paper,oneside,12pt,english]{article}

\usepackage[pdftex]{color,graphicx}

\usepackage[T1]{fontenc}
\usepackage[latin9]{inputenc}
\usepackage{url}

\usepackage{amsmath}
\usepackage{amssymb}

\makeatletter
\usepackage[backref=page]{hyperref}
\hypersetup{plainpages = false,
              breaklinks=true,
              a4paper=true,
              linktocpage,
              pdfauthor={Niko Brummer},
              pdfstartpage=1, 
              pdfstartview=FitH,  
              pdfkeywords={}}

\makeatother

\usepackage{babel}

\def\xvec{\mathbf{x}}

\def\model{\vec{\lambda}}

\begin{document}

\title{What is the `relevant population' in Bayesian forensic inference?}

\author{Niko Br\"{u}mmer and Edward de Villiers \\ AGNITIO Research, South Africa}

\maketitle

\section{Introduction}
\def\pvec{\mathbf{p}}
\def\nvec{\mathbf{n}}
\def\xvec{\mathbf{x}}
\def\qvec{\mathbf{q}}
\def\avec{\mathbf{a}}
In works discussing the Bayesian paradigm for presenting forensic evidence in court, the concept of a `relevant population' is often mentioned, without a clear definition of what is meant, and without recommendations of how to select such populations. This note is to try to better understand this concept. Our analysis is intended to be general enough to be applicable to different forensic technologies and we shall consider both DNA profiling and speaker recognition as examples. 

We shall limit attention to the canonical forensic problem of having to decide whether a given \emph{suspect} and an unknown \emph{perpetrator} of a crime are the same or not. To facilitate this decision, there is given two pieces of evidence: (i) a trace (speech recording, DNA profile, etc.) left behind at the crime scene by the perpetrator; and (ii) a similar sample (recording, DNA profile) obtained from the suspect. 

We base our analysis on the understanding that in its most general form, a relevant population is a \emph{probability distribution} over individuals. As an example, a relevant population for the perpetrator could take the form: the perpetrator is male with probability $90\%$ (or female with probability $10\%$).

We shall discuss the following questions:
\begin{description}
	\item[Q1:] \label{Q1} Who should choose the relevant population, the prosecution or the defence?
	\item[Q2:] \label{Q2} Should the relevant population be associated with the suspect, or the perpetrator, or both?
	\item[Q3:] \label{Q3} Should the relevant population be chosen \emph{before} processing the forensic evidence, or should the forensic evidence be used to make the choice?
\end{description}
We discuss these questions in turn in the following three sections.
 
\section{Q1: Who should choose the relevant population?}
The goal of the forensic inference is to answer the question whether the perpetrator and suspect are in fact the same person or not. The two answers to this question are often termed the \emph{prosecution hypothesis} (they are the same) and the \emph{defence hypothesis} (they are different). This terminology has the advantage that it makes the contents of the propositions immediately clear. But it has the disadvantage that it could lead to the misunderstanding that the finer details of each hypothesis should be respectively formulated by the defence and the prosecution. By `formulating' the hypothesis, we mean making all kinds of hypothesis-dependent auxiliary assumptions, including assumptions about the relevant population. 

We argue below however that to be able to perform a coherent inference, all parties should be in agreement on all assumptions, \emph{except} on the basic question at hand. To avoid this misunderstanding, we therefore prefer to use a more neutral terminology for the two hypotheses:
\begin{itemize}
\item [$H_1$:] The suspect and perpetrator are \emph{one} and the same person.
\item [$H_2$:] The suspect and perpetrator are \emph{two} different people.
\end{itemize}
By inference, we mean reasoning in the face of uncertainty and the Bayesian way of doing this is to compute a \emph{posterior probability distribution} of the form:
\begin{align}
\label{eq:hpost}
P(H_1|E,B) &= 1- P(H_2|E,B) 
\end{align}
where $E$ denotes the evidence subjected to forensic analysis (e.g. DNA profiles, or speech samples) and $B$ denotes background information and assumptions. Note that everything to the right of the `given' symbol, $|$, is the \emph{same} for both posterior probabilities, otherwise we would not get a well-defined probability distribution that sums to 1.

If in contrast, one uses the prosecution/defence hypothesis terminology and additionally, one does not think in terms of the posterior probability distribution, but instead of a likelihood-ratio between these hypotheses, then it is easy to fall into the trap of formulating a \emph{meaningless} likelihood ratio of the form
\begin{align*}
\frac{P(E | H_\text{pros}, B_\text{pros})}
{P(E | H_\text{def}, B_\text{def})}
\end{align*} 
where prosecution and defence assume \emph{different} background information. If the inference is to proceed via a likelihood ratio, it should be of the form:
\begin{align*}
\frac{P(E | H_1, B)}
{P(E | H_2, B)}
\end{align*} 
where $B$ is the \emph{same} in numerator and denominator.
Our answer to 
Q1 is therefore:
\subsection{Conclusion for Q1}
Whoever chooses the background information (which may include specification of a relevant population), should take care that it is the \emph{same} when computing the two likelihoods for the two hypotheses.

\subsection{Discussion}
In principle, it should not matter all that much who chooses the background information. The end result of the inference is a statement of the form:
\begin{itemize}
	\item $B$, the collection of background assumptions.
	\item The evidence, $E$.
	\item The posterior, $P(H_1|E,B)$.
\end{itemize}
In this format, the background information is available for scrutiny. If $B$ is agreed upon, then also $P(H_1|E,B)$ should be agreed upon. 

\subsubsection{Note on posterior vs likeihood ratio}
It is often argued that the forensic scientist should not assign certain types of prior information, so that the whole of $B$ as defined here is not available to him/her. Consequently, the posterior also cannot (and should not) be computed by the forensic scientist. The scientist is then expected to supply whatever parts of the calculation can be done without the unspecified prior information, in such a format that, if the missing prior information were to be supplied, the posterior could be computed in a straight-forward way. The canonical example is to factor the posterior odds as the product of a prior odds and a likelihood-ratio. We shall return to this point later: unfortunately the simple multiplicative formula becomes more complicated if there is significant (posterior) uncertainty about the relevant population.

\section{Q2: Should the relevant population be associated with the suspect, or the perpetrator, or both?}
We are doing probabilistic inference in the face of uncertainty. The primary uncertainty of interest is the identity of the perpetrator. However, in any realistic probability model there will be further unknowns (sometimes called nuisance variables). To do a concrete calculation, we need to assign probability distributions to all unknowns. The choice of relevant population can be understood as conditioning these probability distributions. We shall make this more concrete by way of an example.

\subsection{Example}
\label{sec:example}
Our forensic technology is speaker recognition. The crime occurred in an environment where the population may be characterized in terms of two attributes, namely gender and accent. There are two genders, male and female and there are two accent classes, native and non-native. This partitions the set of speakers of interest into four subsets. 

The crime is the theft of a mobile electronic device which runs a tutorial application to teach non-natives how to speak the native language. All parties agree to condition the inference on the assumption: \emph{the perpetrator is non-native}. This fact is therefore included in the background information, $B$. There is however no further prior information to make a male perpetrator more likely than a female, so all parties agree to assign:
\begin{align}
P(\text{perpetrator is male}|\pi_g) &= P(\text{perpetrator is female}|\pi_g) = \frac{1}{2}
\end{align} 
where we have introduced the label $\pi_g$ for later reference to this prior. Naturally, $\pi_g$ is to be included in $B$.

The stolen device is later recovered, in possession of the suspect, who is arrested. It is discovered that there is a voice recording, denoted $X_r$, on the device, which has been recorded in the course of exercising the tutorial. It is agreed that this recording is the voice of the \emph{perpetrator}, although the suspect claims it is not his voice and that he does not know how the device came to be in his possession.

The suspect is clearly male, so that the fact \emph{the suspect is male} is also included in $B$. 

The suspect has agreed to provide a speech sample, denoted $X_s$, but is otherwise unwilling to give any further information which could characterize him as a native or non-native speaker. Before considering the suspect speech sample, all parties agree to assign the prior: 
\begin{align}
P(\text{suspect is native}|\pi_a)&=P(\text{suspect is non-native}|\pi_a)=\frac{1}{2}
\end{align} 
where again $\pi_a$ is for later reference and is included in $B$. 

Denoting the combined evidence as $E=(X_r,X_s)$, the end goal of the inference in this example is to compute the posterior $P(H_1|E,B)$, which we shall do in section~\ref{sec:Q3} below.

For now, notice that we have effectively chosen \emph{two different} relevant populations. Before analysing the speech samples:
\begin{itemize}
	\item the suspect has been classified as belonging to a population of \emph{native and non-native males},
	\item while the perpetrator has been classified as belonging to a population of \emph{non-native males and females}.
\end{itemize}
It is important to notice that the two populations have a non-empty intersection (non-native males). If the intersection were empty that would have deductively proven $H_2$ to be true, even before the speech samples are analysed.

We can now answer Q2.
\subsection{Conclusion for Q2}
There are \emph{two} relevant populations, one containing the suspect and the other containing the perpetrator. These populations may be different, but if their intersection is empty then $H_2$ is trivially true and we do not need any further analysis of the forensic evidence. 

\subsection{Discussion}
In a very simplistic forensic calculation, the need for choosing a relevant population for the suspect may be unnecessary. This is generally the case when there are no nuisance parameters with unknown values. Examples of nuisance parameters include \emph{hidden variables} in generative models, as well as \emph{unknown model parameters}. 

Consider first the case of DNA profiling where all model parameters are given and where the possibility of profiling errors is negligible, so that the measured profiles of the suspect and perpetrator are assumed to be faithful representations of the true profiles. The perpetrator is unknown, so we still need to assign a prior probability distribution (defined by a relevant population) over possible alternative perpetrators. But the suspect profile is given, reducing the relevant population of the suspect to a set with just one member.

However, as soon as we introduce nuisance variables with unknown values the need for the suspect relevant population resurfaces. We give two examples:
\begin{itemize}
	\item We continue with the DNA example, and we still assume profiling errors to be negligible. But now (as is done in real DNA calculations), we are not given fixed values for allele frequencies in the relevant population. Now new data (such as the suspect's profile) could change the predictive probabilities for observing additional examples of the same profile. As noted above, the forensic analysis is done under the assumption that the suspect is a member of the relevant population from which the perpetrator came. But if this population consists of subsets, we may need to ask to which subset the suspect belongs in order to properly apply the predictive probability updates.
	\item In generative speaker recognition modelling, one supposes that each speaker has a constant, but \emph{unobservable} speaker identity variable. The observed speech is considered to be a noisy version of the speaker identity variable. To test whether two observations come from the same speaker or not, the values of the hidden identity variables have to be inferred. This inference needs priors on the hidden variables and these priors could depend on specification of the relevant population. This would then require specification of the relevant populations for both the perpetrator and the suspect.
\end{itemize}

\section{Q3: Should the forensic evidence be used to facilitate the choice of relevant population?}
\label{sec:Q3}
\def\model{\mathcal{M}}
To answer the final question, let us crank the handle of probability theory to compute the posterior~\eqref{eq:hpost} for the running example, introduced in section~\ref{sec:example}.

We shall use the terminology that $m$ denotes male; $\bar m$ female; $n$ native; and $\bar n$ non-native. Also let $g(s)$ and $g(r)$ denote the suspect and perpetrator genders and $a(s)$ and $a(r)$ the suspect and perpetrator accent classes. For brevity, we also define $c(s)=g(s)a(s)$ to be the population category of the suspect and similarly, $c(r)$ for the perpetrator.

We suppose four probabilistic models (with parameters) are given, where $\model_{ga}$ denotes a model for speech recordings produced by people of gender $g$ and accent class $a$. Also let $\model=(\model_{mn},\model_{\bar mn},\model_{\bar m\bar n},\model_{m\bar n})$ denote the collection of all four models and we include $\model$ as part of $B$. Let $P(X_r,X_s|H_1,\model_{ga})$ denote the joint probability distribution for two recordings of the same speaker;  and let 
\begin{align}
P\bigl(X_r,X_s|H_2,\model,c(r),c(s)\bigr) &= P\bigl(X_r|\model_{c(r)}\bigr)
P\bigl(X_s|\model_{c(s)}\bigr)
\end{align}
denote the joint probability for two recordings of different speakers, where the speaker population classes are given. 

Finally, before we can complete the calculation of the posterior, we still need one more piece of prior information. We have already assigned a prior probability of $\frac{1}{4}$ to the intersection (male, non-native), but given the assumption that the evidence lies in this intersection, we still need a prior probability for $H_1$. We define:
\begin{align}
\pi_h &= P\bigl(H_1|a(s)=\bar n,g(r)=m\bigr)
\end{align}
Now we pile everything into $B$
\begin{align}
B &= \bigl(\pi_h,\pi_g,\pi_a,g(s)=m,a(r)=\bar n,\model\bigr)
\end{align}
and compute the final answer to the whole inference as the posterior:
\begin{align}
\label{eq:post3}
P_f &= P(H_1|X_r,X_s,B) = P_a P_g P_h 
\end{align}
where
\begin{align} 
P_a 
&= P\bigl(a(s)=\bar n | X_s,\model_{mn},\model_{m\bar n},\pi_a \bigr) \\
P_g 
&= P\bigl(g(r)=m | X_r,\model_{m\bar n},\model_{\bar m\bar n},\pi_g \bigr) 
\end{align}
and
\begin{align}
P_h 
&= P\bigl(H_1|X_r,X_s,\model_{m\bar n},\pi_h \bigr)
\end{align}
We shall give further details of how to compute $P_a,P_b,P_c$ below and also discuss how to translate~\eqref{eq:post3} into likelihood-ratio form. But first, we conveniently use~\eqref{eq:post3} to answer Q3. 

Notice $P_a$ answers the question (after having digested both prior information and evidence) whether the suspect accent is non-native. Similarly, $P_g$ answers the question whether the perpetrator was male. In both of these calculations, the prior and the speech evidence were used in their proper places. The answer to Q3 is therefore clearly \emph{yes, the evidence should be used}. Of course, this was a specific example and the details of the calculation will vary from case to case.

\subsection{Conclusion for Q3}
Both prior and evidence are relevant to inferring sub-population membership of the suspect and perpetrator. Expressing everything in terms of the posterior for the primary question shows how to do these calculations and how to combine their results.

\subsection{Discussion}
\def\odds{\mathcal{O}}
For a more careful motivation and understanding of our calculations above, we continue with our running example and give more details of the calculations. In particular, we show how to formulate the whole calculation in terms of likelihood-ratios, so that the analysis of the evidence can be made independent of the prior information.  

We denote our prior for suspect accent as $\pi_a=P\bigl(a(s)=\bar n|\pi_a\bigr)$ and the prior for perpetrator gender as $\pi_g=P\bigl(g(r)=m|\pi_g\bigr)$. We can now compute each of the three posteriors, $P_a,P_g$ and $P_h$, conveniently in \emph{odds-against}\footnote{We could instead work with the reciprocals, \emph{odds for}, but that gives more complex formulas in this problem.} form:
\begin{align}
\odds_a' &= \frac{1-P_a}{P_a} \\
&= \frac{1-\pi_a}{\pi_a} 
\times \frac
{P\bigl(X_s|\model_{mn}\bigr)}
{P\bigl(X_s|\model_{m\bar n}\bigr)} \\
&= \odds_a \times R_a
\end{align}
Where $\odds'_a$ and $\odds_a$ are respectively the posterior and prior odds \emph{against} the proposition that the suspect accent is non-native and $R_a$ is the likelihood ratio for native against non-native males. Note $P_a$ can be recovered from the odds as:
\begin{align}
P_a &= \frac{1}{1+\odds_a'}
\end{align}
To get a feel for odds-against, notice that if $\odds_a'\ll1$, then $P_a\approx1-\odds_a'\approx1$; if $\odds_a'=1$, then $P_a=\frac{1}{2}$; and if $\odds_a'\gg1$, then $P_a\approx\frac{1}{\odds_a}\ll1$. 

In a similar way, we get the odds against a male perpetrator
\begin{align}%
\odds_g'&= \frac{1-P_g}{P_g} \\
&= \frac{1-\pi_g}{\pi_g} \times
\frac{P\bigl(X_r|\model_{\bar m\bar n}\bigr)}
{P\bigl(X_r|\model_{m\bar n}\bigr)} \\
&=\odds_g \times R_g
\end{align}
and the conditional odds against $H_1$
\begin{align}
\odds_h' &= \frac{1-P_h}{P_h} \\
&= \frac{1-\pi_h}{\pi_h} \times
\frac
{P\bigl(X_r|\model_{m\bar n}\bigr)P\bigl(X_s|\model_{m\bar n}\bigr)} 
{P\bigl(X_r,X_s|H_1,\model_{m\bar n}\bigr)} \\
&= \odds_h \times R_h
\end{align}
We are now in a position to divide the responsibilities: 
\begin{itemize}
	\item The forensic scientist, who analyses the evidence, computes the likelihood-ratios $R_a,R_g$ and $R_g$, independently of the numerical values of the priors $\pi_a,\pi_g$ and $\pi_h$. The likelihood-ratios are reported to the court.
	\item The court (judge/jury) could then theoretically assign numerical values to the priors (in odds-against form) and apply the three multiplicative formulas to obtain posterior odds against. 
\end{itemize}
How should the court now proceed to make a final decision? Each of $\odds_a'$ and $\odds_g'$ could be considered independently and if any of them is too large, then $H_1$ can be rejected on the grounds that there is a reasonable doubt that the relevant populations  intersect. Otherwise, if both are sufficiently small, then $\odds_h'$ can be considered for final thresholding against some `reasonable doubt threshold'. 

In summary, the court could make three independent decisions: if any of the three questions casts reasonable (posterior) doubt, then the court finds for $H_2$. 

\subsubsection{The additive odds-against formula}
However, if the court is just slightly more mathematically adventurous, then a very simple approximation could combine the three steps into a single test. Defining the final posterior odds against $H_1$ to be $\odds_f'=\frac{1-P_f}{P_f}$, where from~\eqref{eq:post3} we have $P_f=P(H_1|E,B)=P_aP_gP_h$, we can compute:
\begin{align}
\frac{1}{1+\odds_f'} &= \frac{1}{1+\odds_a'}\frac{1}{1+\odds_g'}\frac{1}{1+\odds_h'} 
\end{align}
or
\begin{align}
\odds_f' &= (1+\odds_a')(1+\odds_g')(1+\odds_h')-1 \\
&=  \odds_a'+\odds_g'+\odds_h' +\epsilon
\end{align}
where $\epsilon$ is a sum of products of the odds. \emph{If} all of the odds are small, then each of these products will be very small compared to the sum of the odds and $\epsilon$ can be safely ignored. Now consider\footnote{In \emph{Probability Theory: The Logic of Science}, E.T. Jaynes suggests in a section entitled `Bayesian jurisprudence' that judges might sleep soundly if they take $\theta=\frac{1}{10~000}$.} some \emph{reasonable doubt threshold}, $\theta\ll1$, on the final odds: The court finds for hypothesis $H_1$ only if $\odds_f'<\theta$. 

To help to understand this threshold, notice that if $\odds_f'<\theta\ll1$, then $\odds_f'\approx1-P_f=P(H_2|E,B)$. If we make this threshold small, we are saying that we require the posterior probability for innocence to be small. 

With the above approximation, the test is now performed as: 
\begin{align}
\odds_f' \approx \odds_a'+\odds_g'+\odds_h' < \theta
\end{align}
and notice that if the sum is below the threshold, then also each of $\odds_a',\odds_g',\odds_h'<\theta$, so that $\epsilon<3\theta^2+\theta^3\ll\theta$ and can indeed be ignored. 

In summary, the court has to do two things:
\begin{itemize}
	\item Multiply each of the three likelihood-ratios by prior odds to get posterior odds.
	\item Find for $H_1$ only if the sum of the posterior odds is below a small \emph{reasonable doubt threshold}. 
\end{itemize}

\subsubsection{The hidden variables}
This section gives further clarification of the interaction between hidden variables and the relevant population. The following form of generative model has been shown to work well in speaker recognition. We suppose that every person has a hidden identity variable (denoted $Y$), which remains constant. Observations of the speech (denoted $X$) of an individual will of course vary and we model this variation as $P(X|Y,N)$, where  $N$ is some probabilistic model for this variation. To complete the picture, we need priors for these hidden variables and this is where the relevant population is needed: In this case, the relevant population conditions the prior as $P(Y|\model_{m\bar n})$. The likelihoods mentioned above can now be more explicitly specified as:
\begin{align}
P(X|\model_{m\bar n}) &= \int_\mathcal{Y} P(X|Y,N) P(Y|\model_{m\bar n}) dY
\end{align}
and
\begin{align}
P(X_r,X_s|H_1,\model_{m\bar n}) &= \int_\mathcal{Y} P(X_r|Y,N) P(X_s|Y,N) P(Y|\model_{m\bar n}) dY
\end{align}
where $\mathcal{Y}$ is the support of $P(Y|\model_{m\bar n})$.

\subsubsection{Model subdivision}
Still in the context of our running example, a natural question to ask would be whether it was really necessary to subdivide our population into four categories. Could we not simply use a single model for the whole population and get a much simpler recipe? On the other hand, what about further subdivision, for example into age categories and so on? 

Such questions cannot be answered without much experience with different models and their interaction with real data. However, we can make some general comments.

What we have essentially done in this example was to represent the distribution for the hidden variables as a mixture model, with four components. Each of the components could for example be multivariate Gaussian. Now a mixture of Gaussians is a different model than just one multivariate Gaussian. Different models lead to different speaker recognizer accuracies. The designer of a speaker recognition system will in general try different model configurations and choose the one which gives the best accuracy. 

Of course, subdivision of the model also supposes that we have enough development data of each category and it is preferable if the data is labelled according to category. In DNA profiling there are similar considerations, definitions of sub-populations depend on the availability of DNA profile databases. In short, the data available to develop a recognizer will be an important consideration in deciding whether to subdivide the model.

An advantage of a subdivided model is that we can make use of more detailed prior information. In our example, we were able to exclude one of the four categories (native female). This effectively makes the prior perplexity in the problem less and this could be expected to give better accuracy.  When pertinent prior information from non-speech sources is available, then a subdivided model can allow this information to be combined with the information that can be extracted from the speech. 

In contrast, the calculation with a monolithic model would have no inputs for suspect and perpetrator gender and accent. The implicit values for these parameters could be expected to more or less agree with the proportions in the development data. But if the resulting speaker recognition model gave good accuracy anyway, then we would expect a good result regardless, because speakers from different population groups (e.g. male/female) should usually be quite easy to tell apart. If the model finds that a pair of different speakers (who happen to be male and female) are most likely different, then the extra information about their genders is unlikely to add much value.

In summary, model (population) subdivision depends on may things and the level and nature of the subdivision is part of an optimization process to improve accuracy.

\subsection{Generalization}
In our running example we had strong prior information that allowed us to make hard decisions about suspect gender and perpetrator accent, before analysing the speech. If these facts were also a-priori uncertain, then the relevant populations for suspect and perpetrator would both have been mixtures of four models, although the mixture weights could in general be different for suspect and perpetrator.

Consider a general case where there are $K$ categories in the population (in our example we had $K=4$). For each category, $C_k$, we have a model $\model_k$. Suppose also we have a prior probability distribution, $\pi_s=(\pi_{s1},\pi_{s2},\ldots,\pi_{sK})$ for the category which contains the suspect and similarly a prior distribution $\pi_r$ for the perpetrator's category. This would give the more general formula for the final posterior as:
\begin{align}
P_f &= P(H_1|X_s,X_r,B)\\
& = \sum_{k=1}^K P(s\in C_k|X_s,\model,\pi_s) P(r\in C_k|X_r,\model,\pi_r) P(H_1|X_s,X_r,\model_k,\pi_{hk})
\end{align}
where $\pi_{hk}$ is the prior for $H_1$, given that suspect and perpetrator are both in category $k$.

\section{Conclusion}
We urge researchers in Bayesian forensic inference to be careful about jumping straight to likelihood-ratio calculation. Considering the posterior as the final idealized goal of the inference can be a valuable tool in making the whole inference process sound.

\end{document}